# Parametric study of cycle modulation in laser driven ion beams and acceleration field retrieval at femtosecond time scale

M. Schnürer[1], J. Braenzel[1], A. Lübcke[1] and A.A. Andreev[1,2,3]

[1]Max-Born-Institut, Max-Born-Str. 2a, 12489 Berlin, Germany
[2]Extreme Light Infrastructure - Attosecond Light Pulse Source (ELI-ALPS), Dugonicster 13, H-6720 Szeged, Hungary
[3]St. Petersburg State University, University emb., 7/9, 199034 St. Petersburg, Russia

High frequency modulations appearing in the kinetic energy distribution of laser accelerated ions are proposed for retrieving the acceleration field dynamics at femtosecond time scale. Such an approach becomes possible if the laser-cycling field modulates the particle density in the ion spectra and produces quasi time stamps for analysis. We investigate target and laser parameters determining this effect and discuss the dependencies of the observed modulation. Our findings refine a basic mechanism, the Target Normal Sheath Acceleration, where an intense and ultrafast laser pulse produces a very strong electrical field at a plasma-vacuum interface. The field decays rapidly due to energy dissipation and forms a characteristic spectrum of fast ions streaming away from the interface. We show that the derived decay function of the field is in accordance with model predictions of the accelerating field structure. Our findings are supported by 2-dimensional particle-in-cell simulations. The knowledge of the femtosecond field dynamics helps to rerate optimization strategies for laser ion acceleration.

I. INTRODUCTION

Possible applications of laser accelerated ion beams (cf. e.g. [1-5] and references therein) need tailored energy distribution functions (EDF) of the ions. This EDF is coupled to the dynamics of the field which accelerates the ions. The fundamental mechanisms of laser driven ion acceleration have been studied for nearly two decades. Special attention was drawn to the Target Normal Sheath Acceleration (TNSA) [6] (and references therein) and the Radiation Pressure Acceleration (RPA) [7, 8]. TNSA is, to some extent, a robust process; while RPA yields highest energies of accelerated ions. The underlying dynamics of the acceleration field has been successfully probed at a timescale of picoseconds [9] down to about one hundred femtoseconds [10, 11]. This knowledge is important for benchmarking theoretical models which allow extrapolation of required laser parameters for desired ion beam parameters.

In this work, we suggest to use a modulation phenomenon in laser driven proton beams for tracing the acceleration process at a femtosecond time scale. This allows comparing experimental EDF results with model functions of the acceleration field dynamics. The knowledge of these fast evolving processes is important to understand limitations of optimization efforts in laser ion acceleration. For our analysis we use and study in more detail faint density modulations in the kinetic energy spectrum of TNSA protons [12] which become visible if the laser intensity contrast between the pulse background and the peak is at a level of about $10^{10}$. At such conditions even the cycling action of the laser field within the temporal envelope of the laser pulse imprints onto the acceleration field and the velocity distribution of the accelerated ions. Our interpretation is based on numerical and analytical model calculations [12] which showed that the laser-cycle driven variation of the ponderomotive potential releases hot electron bunches per cycle such that the sheath field in the TNSA-process is modulated which translates to density variation in the accelerated ion bunch. The lightest ion, the proton, is most sensitive to such a kind of field variation. In combination with proton beam modulation we have also observed at some conditions the same modulation effect in the carbon spectra, which is not further discussed here. An exclusive appearance of beam density modulation in low-energy, low-charge-state carbon ions has been discussed very recently [13]. This finding as well as other ion beam modulations [14, 15] has been observed with different characteristics at specific interaction parameters.

First, we will discuss how the observed density modulation in the proton spectrum depends on laser parameters (polarization, temporal contrast) and on target parameters (thickness of the target foil, surface modification via structuring). Second, we use our interpretation of the modulation effect to retrieve the acceleration field strength at femtosecond time scale. This result we compare to model predictions and 2D PIC- simulations describing the decline of the acceleration field. We conclude that the process of energy dissipation and the corresponding decline of the acceleration field strength are related to the gain of kinetic energy during the ion acceleration in the field.

## II. EXPERIMENT

Ion acceleration experiments were performed with the 70 TW Ti:Sapphire laser arm in the former High Field Laboratory at Max Born Institute. The central wavelength of the laser was about 800 nm, the pulse duration was ~35 fs and the maximum laser energy on target reached up to ~2 J. Focusing the laser pulse with a f/2.5 off-axis parabolic (OAP) mirror to spots with FWHM of ~ 4 µm realized peak intensities of up to ~ $10^{20}$ W/cm$^2$. The incident angle of the laser beam is between 0.1 and 1 degree in order to avoid exact back-reflection. In comparison with our first work [12], the second arm of the former two beam High Field Laser system (other preamplifier architecture – multi-pass instead of regenerative) was used, allowing higher laser intensities and the possibility of switching to ultra-high temporal contrast with a plasma mirror. Furthermore, a different Thomson-spectrometer configuration with an electric field to separate different ion species was employed. The temporal intensity contrast of the laser pulse was realized by a cross polarized wave generation (XPW) front end yielding a peak-to-amplified spontaneous emission (ASE) contrast of better than $10^{10}$ (high contrast - HC). For driving ultra-thin targets the contrast was further be enhanced by a double plasma mirror (DPM) and reaches values better than $10^{14}$ (ultra-high contrast - UHC). Targets are 5 µm thick titanium foils or 30 nm thick plastic (poly-vinyl-formal PVF, $C_5H_7O_2$, density 1.23 g/cm$^3$) foils [16]. Kinetic energies of accelerated ions are measured with a Thomson Mass Spectrometer using a slit ((50 – 340) µm width) as an entrance aperture. Recording the momentum dispersed ion traces with such an entrance slit allows discriminating faint density modulations against noise. This is a mandatory prerequisite.

The entrance slit for the Thomson-mass spectrometer is oriented parallel to the magnetic and the electric field (both fields have the same orientation). Therefore ions with the same kinetic energy and the same charge to mass ratio appear at the detector plane as a line being perpendicular to the magnetic deflection direction and in line with the electric field direction. The bending of the traces (Photo of the parabolic traces in Fig. 3) is caused by the electric field. The width of the traces is determined by the length of the slit. Therefore, our observation is a proof that the density maxima of the signal (lines) and their orientation corresponds to ions (protons) with the same energy.

An imaging Hamamatsu MCP, 100 mm in diameter, detected the ions and the visible Thomson parabolas on the fluorescence screen were recorded with a CCD camera system. The detector response was cross checked with alpha-particles from a radioactive $Am^{241}$ source. In most experiments, p-polarized laser pulses irradiated the target foils close to normal incidence.

For some experiments, in order to generate circularly polarized pulses, a 200 μm thick mica λ/4 plate was inserted in front of the second to last turning mirror delivering the beam to the focusing OAP mirror. The polarization setting with the λ/4 wave plate was aligned with a Glan-Thompson prism. In order to visualize the density modulations Figs. 1 and 2 display proton raw-signals of a detector pixel without considering the binned energy interval of the pixel, in Fig. 4 the proton number is calculated per binned energy interval for comparison.

### III. PARAMETRIC STUDY OF MODULATION IN THE ENERGY DISTRIBUTION OF LASER ACCELERATED PROTONS

In past proton imaging experiments [17], using a slit-spectrometer, the temporal pulse contrast was at a level of $10^6$ - $10^7$ and this did not allow observing the modulation effect. For the first time [12] we could detect high frequency modulation of the proton EDF when an XPW-frontend was set into operation and consequently the temporal intensity contrast was enhanced by nearly 3 orders of magnitude (HC).

Within the experiments described in reference [12] we inserted thin Al-foils into the ion beam path in front of the spectrometer slit and repeated the experiment without any other changes. The scattering in the filter foil changed the longitudinal emittance of the proton beam and the weak modulation feature disappeared. This is a strong proof to exclude any instrumental artifact as e.g. structures of the imaging detector or high frequency modulation of the deflecting fields inside the spectrometer. Within the course of this work we describe parametric target and laser dependencies of the modulation effect.

In order to substantiate the modulation in laser driven ion acceleration further, it is useful to identify additional conditions which suppress the effect. Changing the laser polarization from linear to circular for a normally incident light wave at a plane surface is one obvious strategy because the electron trajectories in the light field are different and the forces accelerating the electrons are different, too. (In case of circular polarization the vector product between the electron velocity and the B-field of the electromagnetic wave vanishes, at linear polarization it does not.) This in turn changes the oscillatory motion of the electrons and in specific also the hot electron population which is responsible for the TNSA-mechanism. But, as visible in Fig.1, the experiment with a 5 μm thick titanium foil at HC-condition does not show any significant difference between linear and circular polarization.

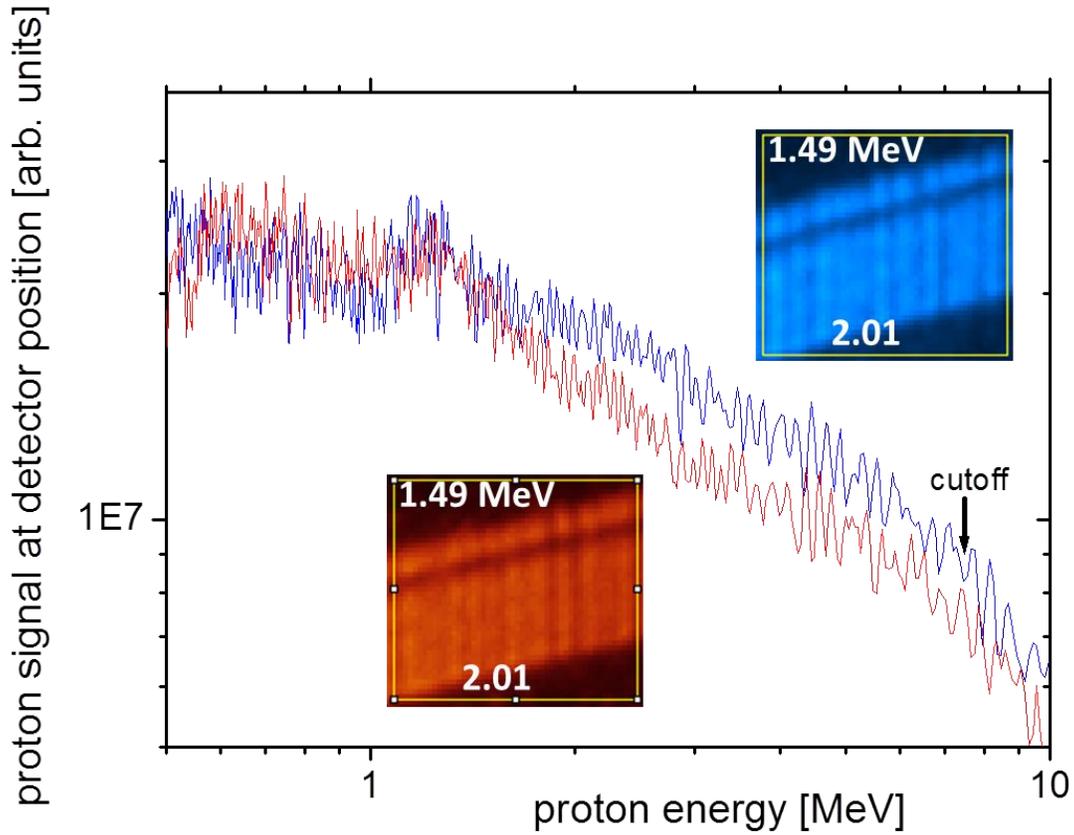

Fig.1 Fast protons produced with laser irradiated 5 μm Ti-foils using linear (blue line) and circular (red line) laser polarization (HC, parameters cf. text). The proton signals are pixel counts at pixel position (at calculated energy position) of the detection system. (insert: contrast enhanced parts of the recorded proton traces)

In order to substantiate this observation we picked six successive laser shots (p1 to p6) from one experimental run with the same laser energy irradiation and focusing (laser intensity at FWHM ~ 7 x 10$^{19}$ W/cm$^2$, high contrast) and target conditions (5 μm thick titanium foil) in Fig.2. We applied different spectrometer slit widths and a laser polarization as described in the caption of Fig.2.

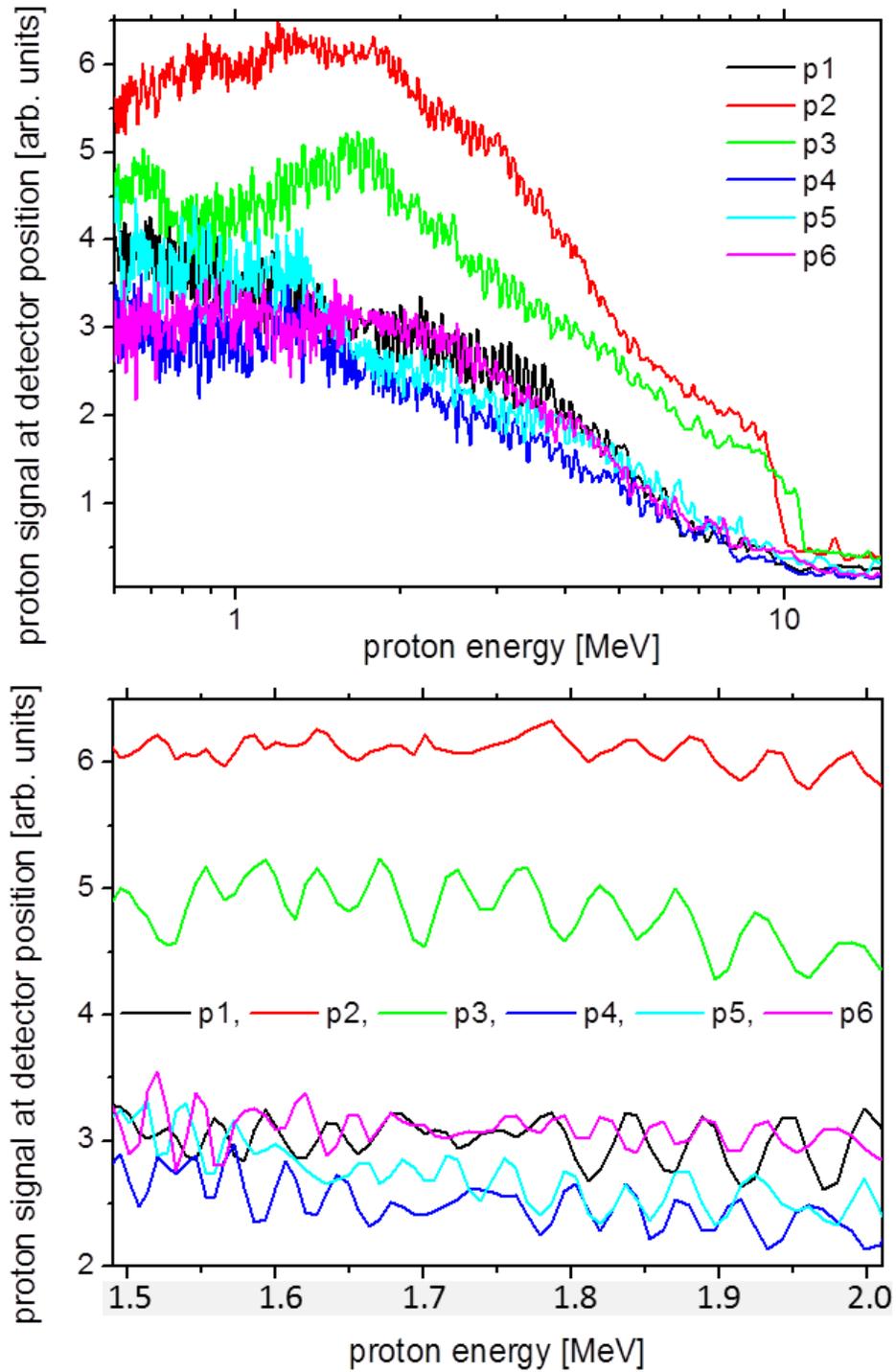

Fig.2 Fast protons produced with 7 x 10[19] W/cm[2] laser irradiated 5 μm Ti-foils (lower graphic shows parts of the traces with zoomed-in energy interval similar to inserts in Fig. 1): linear polarization applied for p1-black, p5-cyan, p6-purple and circular polarization applied for p2-red, p3-green, p4-blue; spectrometer slit width were 340 μm for p1 and p2, 240 μm for p3, 50 μm for p4, p5 and p6;

In all shots the proton cutoff energy reaches 10 MeV. Although our available data are not backed with a statistics built on hundreds of shots the presented successive ones in Fig.2 demonstrate that the observed modulation phenomenon is reproducible if laser and target parameter allow the phenomenon to occur. Shot to shot variation of signal strength and spectral shape are apparent. This becomes visible if the signals are recorded with different slit width. While smaller slits lead to lower signals, the signal reduction did not follow exactly the reduced slit width. The used imaging system for the phosphorous screen had a resolution of 0.108 mm per CCD-camera pixel. The MCP has a channel diameter and channel pitch of about 25 μm and 31 μm, respectively. The width of the density maxima is about 2 – 3 CCD-pixel with the 50 μm slit and 3 – 4 CCD-pixel in case of the 240 μm slit. The slit itself was at a distance of 40 cm from the target and the MCP-detector was 100 cm located from the slit. The center of the magnet and the electric field plates was about 35 cm behind the slit.

For the smallest slit width the signal fades already around the cutoff. The modulation phenomenon is clearly reproduced in all shots and does not show a qualitative difference between linear (LP) and circular (CP) polarization. A similar number of peaks (11 – 14) appear in the selected energy interval. The peak position fluctuates as expected for a non-phase locked multi optical cycle femtosecond laser pulse. In different experiments with ultra-thin target foils we applied also this switching technique between LP- and CP-pulses. We investigated High Harmonic Generation in laser transmission direction and observed emission only in case of LP-pulses. This clearly shows that our technique of CP pulse generation does not leave a strong LP background.

Therefore, we exclude a systematic experimental error. Instead, we will argue in the following, that at the given laser intensities the plane of critical plasma density is bent. The magnitude of the bending angle $\Theta_{sf}$ can be approximated with the analytical expression: *$<tan(\Theta_{sf})> \approx c (I^{18}_L)^{0.5} / (\omega_{pe} w_L)$*, $c$ – speed of light, $I^{18}_L$ – laser intensity as a multiple of $10^{18}$ W/cm$^2$, $\omega_{pe}$ - plasma frequency of electrons and $w_L$ - laser beam waist. The equation is derived for a gaussian laser beam profile assuming balance of the ponderomotive pressure by the created ambipolar field at the position of the critical electron density [18]. With our parameter set we obtain on the order of $\Theta_{sf} \sim$ 10°, which alters already the interaction significantly. Due to the bent critical surface quite different electron trajectories and acting forces are possible such that the simple assumption of parallel electron velocity and B-field vectors does not hold anymore. In comparison with our experimental results also in theoretical studies modulations are visible with CP-pulses (cf. PIC-simulation visualized in Fig. 2, 3, 4 of a very recent paper [19]). From our experimental results alone we cannot conclude on modulation period. Our analysis in [12] suggests a period of half a laser cycle due to the ponderomotive force in case of LP-pulses. The PIC-simulations for CP-pulses in ref. [19] suggest a modulation close to a laser cycle in a certain parameter range. We did not observe a clear difference between LP- and CP- pulses in our experiments. We have to keep this question open.

These findings point to other effects and also suggest that more parametric experimental studies and respective high resolution simulations are necessary to resolve the detailed plasma dynamics at the temporal scale of the driving laser cycle.

The density of the electron layer at the target surface and the creation of the hot electron population is a function of the temporal contrast. Therefore it is interesting to study how the modulation effect changes in respect to the temporal contrast of the laser pulse. Applying UHC laser pulses to 5 µm thick Ti-foils leads to visible though much weaker modulations. If now an ultrathin (here a 30 nm plastic foil) target is exposed with UHC pulses the modulation effect vanishes. This comparison is illustrated in Fig.3. The visibility or occurrence of modulations is not an effect of the signal strength as it is visible in the equal measure color coded photos and profile scans in Fig.3.

In case of ultra-thin foils, which have low mass, the whole target starts to move during the irradiation and, probably more important, the illuminated target area is bent which changes the sheath field geometry at the target rear. This is different in case of thick targets where bending (or hole boring) influences the illuminated target front side only. Bending of the whole target foil might hinder small variations of the sheath field to become effective in an ion signal that is always obtained by integration over a finite emission area and volume.

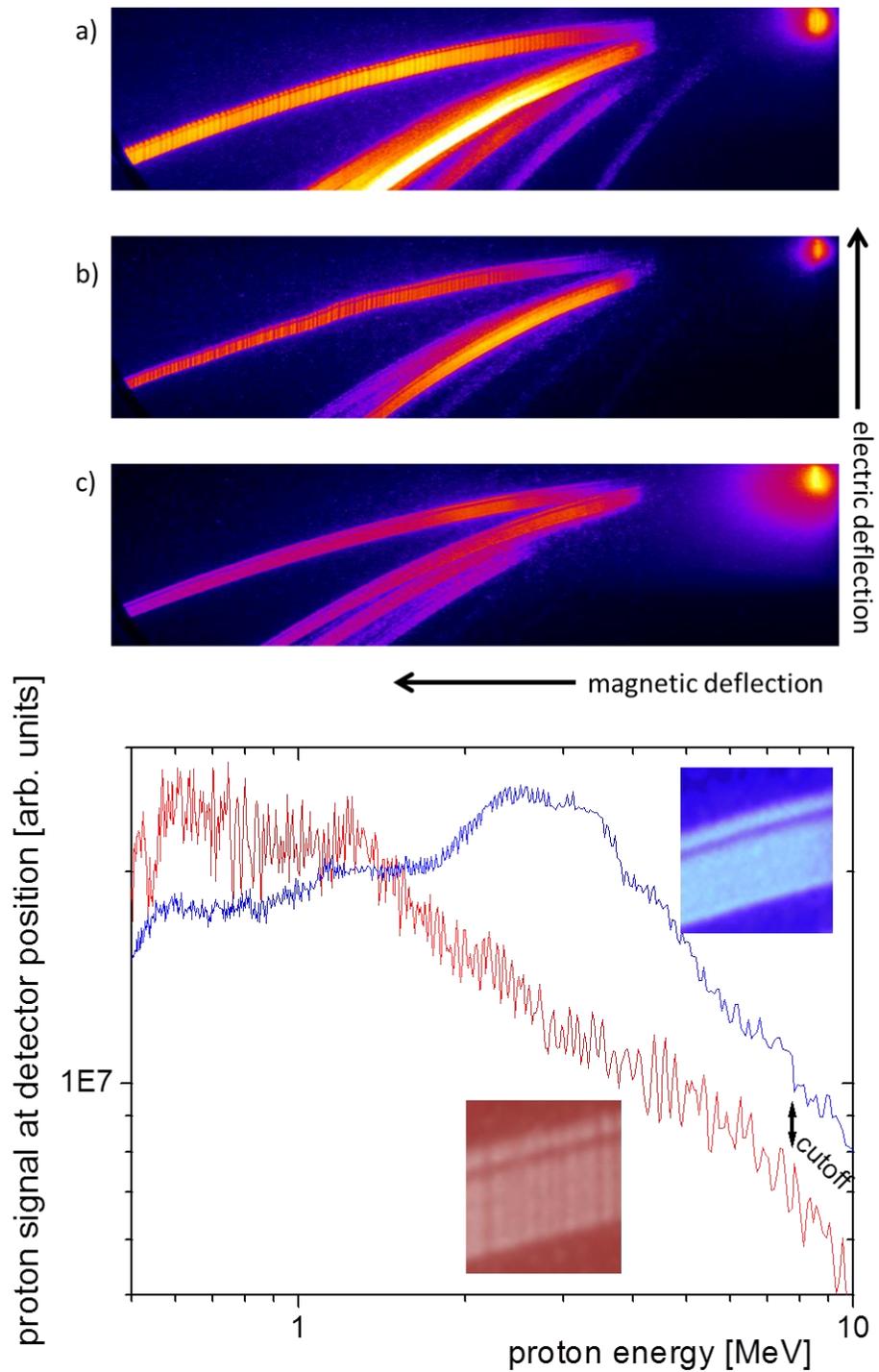

Fig.3 upper part: recorded detector images of the ion spectra (parabolic traces) in same false color scaling - a) 5 µm Ti-foil, HC, circular polarization, spectrometer slit 240 µm, b) 5 µm Ti-foil, HC, linear polarization, spectrometer slit 50 µm and c) 30 nm plastic foil (UHC, linear polarization, parameters cf. text), spectrometer slit 50 µm; lower part – scans of proton traces from pictures a) and c): 5 µm Ti-foil (HC, circular polarization, red line, cf. Fig.1), 30 nm plastic foil (UHC, linear polarization, parameters cf. text) (blue line), insert – magnified part of signal traces;The proton signals are pixel counts at pixel position (at calculated energy position) of the detection system.

In addition, we studied how a target surface structure can influence fast ion generation within the TNSA regime. Laser induced periodic surface structures (LIPSS) were created in situ and a different parametric study was performed [20]. Here, the laser irradiated surface is modified to enhance the absorption of the high intensity laser pulse which drives the ion beam from the plane rear surface of the target. The use of the slit-spectrograph configuration revealed that the surface structure introduces observable smearing and bending to the regular modulation structure of the proton beam (cf. inserted pictures in Fig.4). The inserts in Fig.4 show that a slight bending of the line-features in the trace appeared. In this case, the used slit was longer and the spatial dependence of the magnetic field inside the U-type magnet becomes visible. Another proof can be found in the change of the modulation appearance when structured targets were used. All this demonstrates that the observed phenomenon is not an artifact but shows dependencies on target and laser parameter.

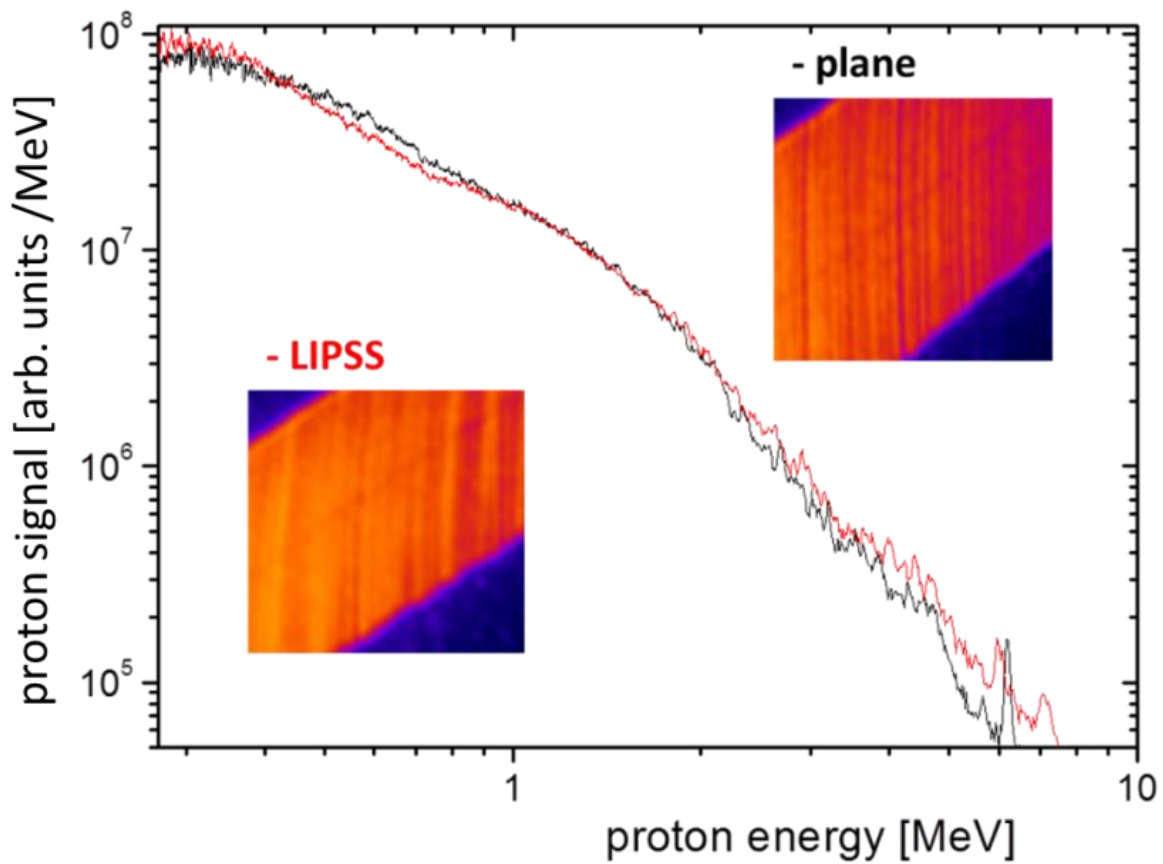

Fig.4 Appearance of density modulation in fast proton bunches produced with HC laser pulses irradiating plane or LIPSS Ti-foils (parameter cf. text). The non-calibrated proton signal is evaluated with energy binning. (insert: contrast enhanced parts of the recorded proton traces)

This is quite understandable because the spatial scale of the created surface structure is on the range of several 100 nm which translates to illumination variation at the order of one femtosecond. This is close to the field modulation cycle of about 1.35 fs (half of the laser field period). Fig.4 shows the comparison between proton spectra obtained with a structured and a non-structured target surface of 5 μm thick Ti-foils. Here, the proton numbers are evaluated in respect to the detected energy intervals. The signal shows a variation over three orders of magnitude and thus the density modulations, which are at a few percent level of the total signal only, are hardly visible in the graphs without zooming in. Comparative studies between structured and non-structured targets were performed [20]. Also these results gave motivation to investigate the time dependence of the hot electron distribution further which determines ion acceleration and X-ray emission.

## IV. ACCOMPANYING PARTICLE-IN-CELL SIMULATION

Particle-in-cell (PIC) simulations were performed at a laser intensity of 5 x $10^{19}$ W/cm$^2$ using the two-dimensional modified PSC code [23]. The simulation box (15000 × 5000 cells) is 30 × 25 µm$^2$ large with step sizes of 2 nm in longitudinal and 5 nm in transversal direction and 30 particles per cell are included. The time step is 2 nm/$c$. The interaction is collisionless and the simulation uses periodic boundary conditions. A high contrast laser pulse (35 fs, 4 µm spot size, super-Gaussian/Gaussian in space/time, interacts with a Ti target (density 6 × $10^{22}$ cm$^{-3}$). An average ionization state <$Z$> = 10 is evaluated from the Ammossov-Delone-Krainov (ADK) model [24, 25]. The target thickness is restricted to 1 µm due to numerical reasons. The simulation yielded a maximum sheath field of 0.55 $E_L$ ($E_L$ – electrical field strength of the laser pulse. This corresponds to 20 x $10^{12}$ V/m using $I_L$ = 5 x $10^{19}$ W/cm$^2$ and $I_L = \varepsilon_0 c E_L^2$. The field strengths obtained from theory and deduced from experiment agree quite well (cf. Fig.6).

We complement the results shown in [12] with Fig.5 which shows the generation and evolution of electron bunches at every laser half cycle in the simulation. These bunches modulate the accelerating sheath field [26] and modulation of the density in the emitted proton beam becomes visible [12].

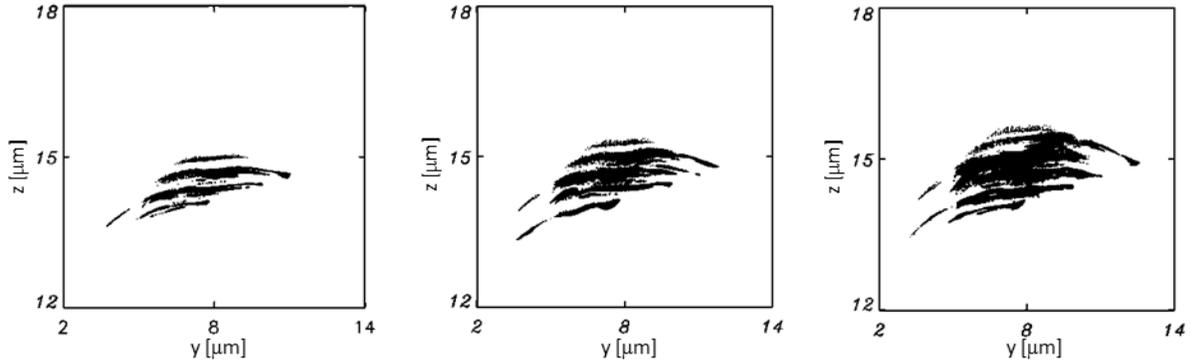

Fig.5 2D-PIC-simulation of electron density evolution at time instants 21.3 fs, 22.65 fs, 23.9 fs (from left to right) intial target position at z= 13.5 µm, laser propagation in z-direction;

## V. SIMPLE APPROACH TO RETRIEVE THE DECAY CHARACTERISTIC OF THE ACCELERATION FIELD

As briefly mentioned in our previous work [12] one can use the recorded modulation maxima ($E_{p\_max,i}$) or minima in the proton EDF in order to estimate the acceleration ($a$) of protons (mass – $m_p$) with a distinct kinetic energy. In the following, in the data interpretation we restrict to a simplified description of the acceleration scenario. It neglects spatial dependencies of the acceleration field and it assumes that different ion energies can be described with a different effective acceleration time in a temporally varying field. We will show that this data interpretation leads to a prediction of the temporal decay of the field which is in accordance to PIC-simulation and suggestions of analytical models.

Every laser half-cycle $\tau_p/2 \sim 1.35$ fs, which is the modulation period of the ponderomotive potential, a maximum/minimum density variation in the proton EDF is formed. The modulation gives a time stamp for the elapsing acceleration time of the ions in an electrical field with decreasing strength. We assume that the acceleration time is $i\ \tau_p/2$. Thus, considering $i$ – acceleration cycles, we can approximate the ion (proton) velocities $v_{p,i}$ with:

$$v_{p,I} - v_{p,i-1} = \Sigma^{i}_{n} a_n\ \tau_p/2\ -\ \Sigma^{i-1}_{n} a_n\ \tau_p/2\ =\ a_i\ \tau_p/2 \qquad (1)$$

Doing so we can take the experimentally determined energies of neighbor peaks $i$ and $i+1$ ($W_{p\_max,i}$) for estimating the acceleration $a_i$ for each $i$ :

$$(W_{p\_max,i})^{1/2} - (W_{p\_max,i-1})^{1/2} = (m_p/2)^{1/2}\ (a_i\ \tau_p/2) \qquad (2)$$

The simple formula uses the approximation that the final EDF of the ions is a result of different acceleration times. This gives a value of the acceleration field strength $E_i$

$$E_i = F_i/e_0 = m_p\ a_i/e_0 \qquad (3)$$

($e_0$ – elementary charge) which is plotted in Fig.6. Because we do not know the absolute point in time when the acceleration starts, we chose t=0 for the first visible modulation peak at high kinetic energy. The proton cutoff energy is higher as the energy of the first peak. Between the cutoff and the first visible peak no modulation is visible. Therefore, extrapolated field values concerning the cutoff, as well as the maximum field value in the PIC-simulation (cf. Fig.6) appear artificially with negative time values. The time, which is assigned to the field decay, is then $i\ \tau_p/2$ ($i$ – number of recorded density maxima in the proton trace, $i_{max}$ = 90).

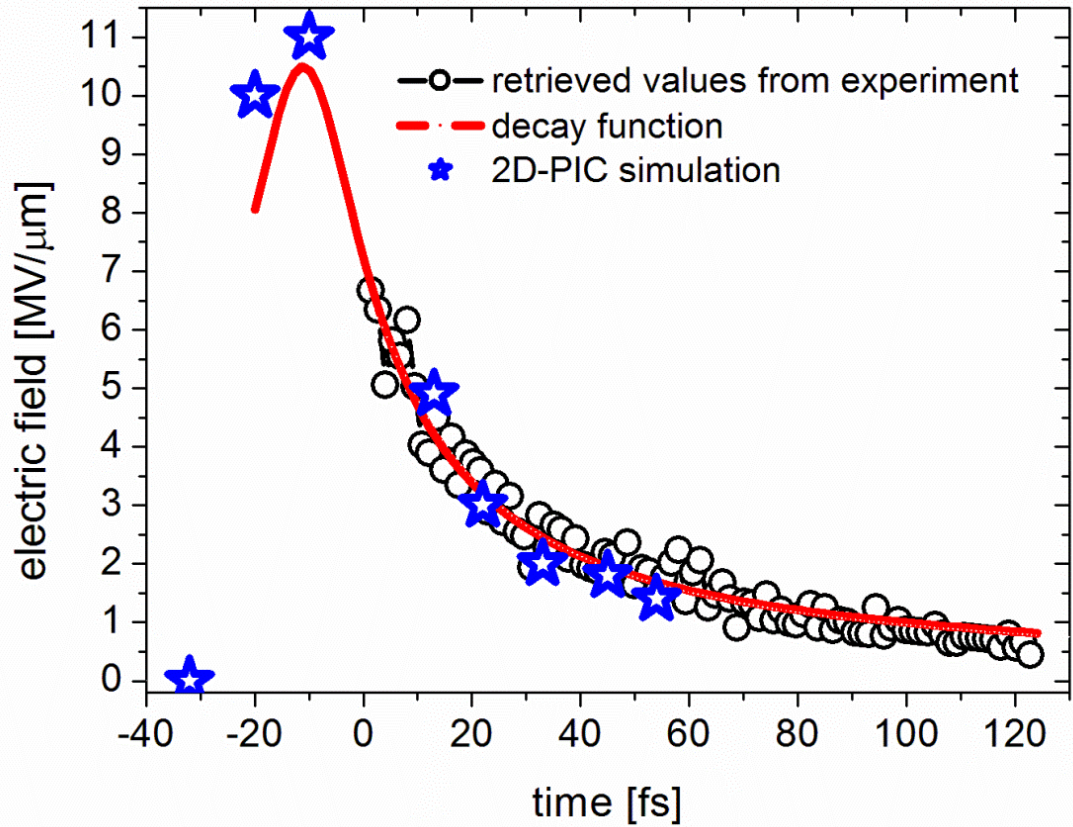

Fig.6 Evaluated decrease of the acceleration field strength (black circles) from the experimentally obtained density modulation in the energy distribution of protons (cf. Fig. 1 – 90 signal maxima of the proton EDF), the red curve represents a nonlinear fit with a decay function (cf. text), blue stars – results from 2D-PIC simulations (cf. text);

The observables are the signal maxima in the proton trace which can be correlated with equation (1) and field strength calculation follows with equations (2) and (3). These retrieved values of the electrical field strength are compared with an analytical function derived in Mora´s work [21] which describes the temporal decay of the field. This function reads as: $E = P_1 (1+P_2((t-t_0) + P_3)^2)^{-1/2}$ (E – electrical field strength, t – time, parameter $P_{1,2,3}$, $t_0$ which are used in the fit procedure). We determine $P_1$ with the following simple linear extrapolation: For the first observed maximum in the proton trace at 6.3 MeV a field value of 6.7 MV/µm is retrieved. This ratio gives for the observed cutoff at 10 MeV a field value of 10.6 MV/µm which we use for $P_1$. The free parameters $P_{2,3}$ and $t_0$ are determined using a nonlinear fitting routine, the obtained red curve is plotted in Fig.5. Here, the used free parameters are not brought into further context with Mora´s model. Apparently, this analytical function describes well the temporal field decay derived from experiment.

But Mora´s model [21] uses an isothermal approach which holds for the duration of the laser pulse. In our case the laser pulse is much shorter than the derived time scale for the acting acceleration field. Therefore a hybrid model approach [22], incorporating also a second adiabatic expansion phase when the laser pulse is off, is more appropriate. Such a situation is simulated in 2D-PIC calculations covering a time window of 90 fs. The result is also plotted in Fig. 4 (blue stars): The data represent the maximum field strength which occurs in the acceleration sheath in target normal direction at different times.

Because the values derived from experiment (black circles) do not allow timing of the laser peak (or pulse switch on), the maximum position of the PIC-data and the decay function curve approximating the experiment were shifted to overlap. The comparison shows reasonable agreement.

Using the same time scale as in Fig.6 we plot the energy of the protons at the appearing density peaks in Fig.7.

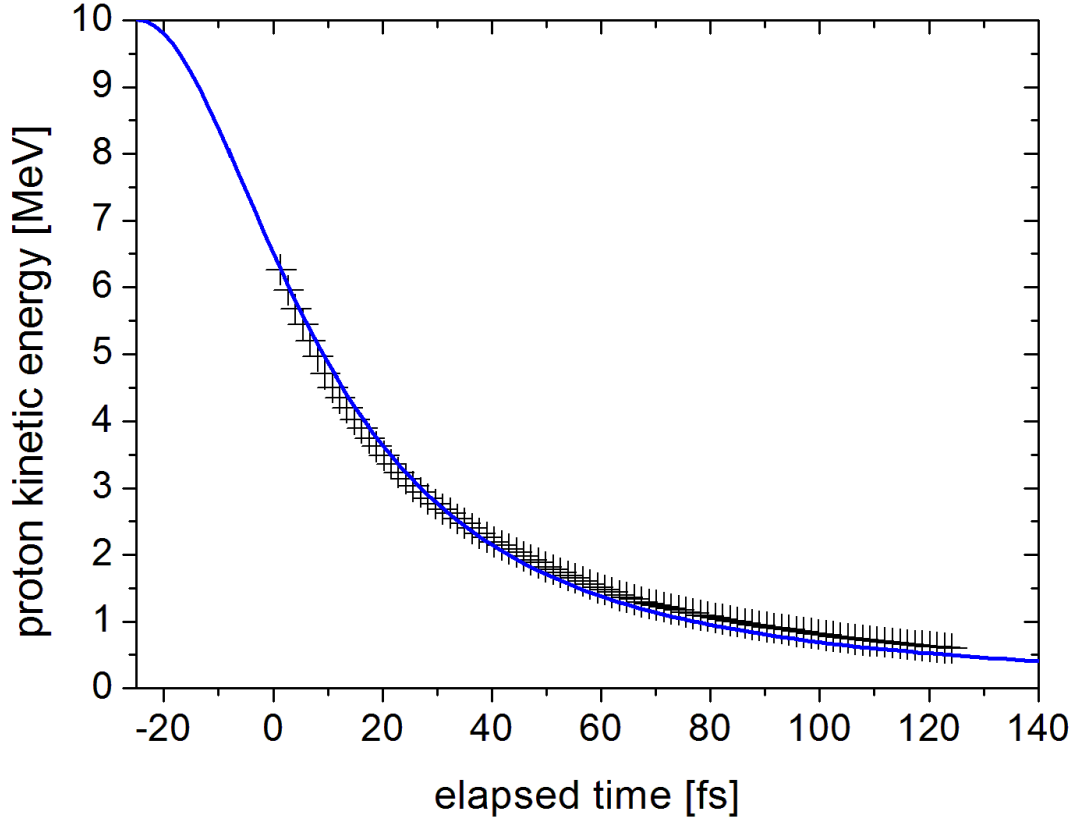

Fig.7 Proton kinetic energy (crosses – experimental data) at observed density maximum plotted at a temporal position concerning the maximum number i and a following elapsed time i *$\tau_p$/2 ~ i *1.35 fs. The blue line represents a model (cf. text) how the proton energy decreases as a function of the hot electron temperature.

In order to also compare the retrieved temporal development of the proton energies $E_p$ with model predictions incorporating adiabatic cooling, the well-known interconnection between the laser-produced hot electron distribution $T_h$ and the TNSA-protons is used. The following set of equations (cf. analytical model in [27]):

$$E_p \approx 2 k_B T_h \quad (4)$$

$$\eta(I) \approx 0.2 + (0.1 + 0.06 \frac{L}{\lambda}) I_{18}^{0.2} \quad (5)$$

$$T_h(I,t) \approx \frac{T_{0h}(\eta(I)I\lambda^2 / 1.37 * 10^{18} \, (W\mu m^2 / cm^2))^{0.5}}{1 + (2t^2 T_{0h} / d^2 m_i)(\eta(I)I\lambda^2 / 1.37 * 10^{18} \, (W\mu m^2 / cm^2))^{0.5}} \quad (6)$$

with $k_B$ – Boltzmann constant, $\eta$ - calculatated laser absorption as a function of laser intensity - $I$, laser wavelength - $\lambda$, scale length - $L$ of the gradient of the electron density at the laser irradiated target surface, $I_{18}$ – the multiple of $10^{18}$ W/cm$^2$, d – target thickness, $m_i$ – ion mass, $T_{oh}$ – model parameter; is used to calculate the proton energy as a function of time.

In order to reproduce the cutoff at $E_p$ = *10 MeV* for an intensity of 5 x $10^{19}$ W/cm$^2$ (FWHM), $\lambda$ = 800 nm and L = 80 nm, (at a time (t + $t_0$) =0 , – $t_0$ fit parameter); we set $T_{0h}$ = *1.58 MeV*. This compares well with the ponderomotive energy of 1.4 MeV, cf. e.g. in [28]. With $t_0$ = *25 fs, $m_i$ = $m_p$ and $d^2$ = 1.1 $\mu m^2$* the model curve in Fig. 6 is calculated. Here, $d^2$ is used as a fit parameter.

We can estimate the energy balance between the stored energy in the sheath field and the total amount of energy released with fast ions via acceleration in the sheath field. This estimation is made on basis of a simple capacitor model for the sheath field. Calculating the energy stored in a capacitor and using an initialized field $E_{sheath\,filed}$ of 10 MV/µm, an area $A_{sheath}$ with a diameter of 10 µm at the target rear (set with a divergent hot electron current from the focal region of the target front), a Debye-length $\lambda_d$ of 1 µm [6] for the charge separation distance in the field, ($\varepsilon_0$ – vacuum dielectric constant), one yields

$$W_{sheath} = \tfrac{1}{2} \varepsilon_0\, A_{sheath}\, \lambda_d\, (E_{sheath\,filed})^2 \quad \approx 35\ mJ\ . \qquad (7)$$

This is at the order of 2% of the incident laser energy, also a consistent value when compared with experimentally determined laser to ion energy conversion efficiencies [29] for our laser and target parameters.

VI.  CONCLUSION AND SUMMARY

One can conclude that the ion acceleration itself is a main channel of energy dissipation for the sheath acceleration structure. This "adiabatic expansion and cooling" is coupled to a fast decline of the acceleration field strength which we could trace with our experimental observation. This behavior sets limits for optimization in laser driven ion acceleration, e.g. if the laser light absorption is enhanced via surface modification or other means. The initially very high values of the acceleration sheath field drop down very fast, nearly 50% in about 30 fs in our case. The fading of the acceleration field lasts much longer. The field decay is connected to the dynamics of the electron EDF. Because different branches of the hot electron distribution develop differently the effect on different types of secondary fast ion and X-ray emission is also different.

Summarizing we can state that we can trace the temporal development of the electrical acceleration field at a femtosecond time scale. The experimentally obtained density modulation in proton spectra and its interpretation give a key to study the temporal field decay which determines the distribution of ion energies. Our results support a fast decay of the acceleration field in the femtosecond time scale in accordance with plasma expansion and ion acceleration models as well as with simulation results. Especially the peak values of the sheath field are much higher as averaged values which are often used for estimations.

Acknowledgement

This project has received funding from the European Union's Horizon 2020 research and innovation programme under grant agreement No. 654148 Laserlab-Europe and by the Deutsche Forschungsgemeinschaft CRC/Transregio 18. We thank D. Sommer for his excellent work on the target foil production and L. Ehrentraut for the laser operation.